# Analytical Fidelity Calculations for Photonic Linear Cluster State Generation

Rohit Prasad,[1,*] Simon Reiß,[2,*] Giora Peniakov,[1] Yorick Reum,[1] Peter van Loock,[2]

Sven Höfling,[1] Tobias Huber-Loyola,[1] and Andreas Theo Pfenning[1]

[1] Julius-Maximilians-Universität Würzburg, Physikalisches Institut, Lehrstuhl für Technische Physik, Am Hubland, 97074 Würzburg, Germany.

[2] Institute of Physics, Johannes-Gutenberg University of Mainz, Staudingerweg 7, 55128 Mainz, Germany.

* Corresponding author(s): Rohit Prasad, Simon Reiß

Tel.: +49-931-31- 81146, +49 6131 39 23628.

E-mail address: rohit.prasad@uni-wuerzburg.de, sreiss@uni-mainz.de

**Abstract.** By precisely timed optical excitation of their spin, optical emitters such as semiconductor quantum dots or atoms can be harnessed as sources of linear photonic cluster states. This significantly reduces the required resource overhead to reach fault-tolerant optical quantum computing. Here, we develop an algorithm that analytically tracks the global density matrix through the process of the protocol for generating linear-cluster states by Lindner and Rudolph. From this we derive a model to calculate the entangling gate fidelity and the state fidelity of the generated linear optical cluster states. Our model factors in various sources of error, such as spin decoherence and the finite excited state lifetime. Additionally, we highlight the presence of partial reinitialization of spin coherence with each photon emission, eliminating the hard limitation of coherence time. Our framework provides valuable insight into the cost-to-improvement trade-offs for device design parameters as well as the identification of optimal working points. For a combined state-of-the-art quantum dot with a spin coherence time of $T_2^* = 535$ ns and an excited state lifetime of $\tau = 23$ ps, we show that a near-unity entangling gate fidelity as well as near-unity state fidelity for 3-photon and 7-photon linear cluster states can be reached.

# 1. Introduction

Quantum computing with photons was shown to be theoretically possible with only linear optics and adaptive measurements already in 2001 [1]. In the same year, an alternative scheme for quantum computing referred to as measurement-based quantum computing was proposed [2]. This scheme uses highly entangled states, called cluster states [3]. Although the original proposal did not have photons in mind, it turns out that using cluster states of photons can be of great advantage for optical measurement-based quantum computing [4].

This is especially true for fusion-based quantum computing [5], which allows for advantageously low error thresholds to reach fault-tolerance by employing fusion operations on few-qubit entangled resource states such as 4-star or 6-ring photonic graph states [6]. Such graph states can be generated from single photons [7–9], nonlinear processes [10], or atom-like transitions [11–14]. From these different ways to generate photonic cluster states, the generation from atom-like transitions, as first proposed by Lindner and Rudolph [15], is perhaps the most efficient, since the creation of the state is deterministic, and the collection efficiency can be engineered to be high. In fact, the required source overhead for fault-tolerant fusion-based quantum computing can be drastically reduced (from millions to a *'handful'* of devices) by employing *Lindner-Rudolph* sources [16,17]. with recent studies tailoring fusion-based photonic quantum computing schemes to *Lindner-Rudolph* sources [18].

Several studies have investigated the impact of imperfections on the cluster state generation protocol, quantifying the effects of distinguishability, higher-order photon emission, and losses. These imperfections have been analyzed in the generation of Greenberger–Horne–Zeilinger (GHZ) states, controlled-phase gates, and Ising and XY interactions, all of which play a role in quantifying entanglement within cluster states. [19,20] Understanding the impact of such errors is crucial for the realization of fault-tolerant quantum computing [21–23]. In this paper, we develop an algorithm that analytically tracks the global density matrix through the process of the Lindner and Rudolph protocol. From this we derive an analytical model to compute the fidelity of the generated cluster states that incorporates real-world parameters, including emitter excited state lifetime, coherence time, spin Larmor precession time, excitation pulse timing, and Landé g-factor. A similar study to investigate sources of error and optimize the experimental parameters to maximize fidelity was recently explored the preparation of graph states using superconducting qubits, specifically fixed-frequency transmon qubits [24]. While an evolutionary approach based on the Markovian master equation has been used to predict the density matrix of linear cluster states (LCS) [11,12], we introduce an analytical model that explicitly captures the effects and correlations of realistic experimental parameters on the fidelity and decomposes the time evolution of the generation protocol into individual, erroneous quantum gate operators. Our model enables efficient optimization of experimental conditions, providing a realistic upper bound on achievable fidelities. Although evolutionary models with continuous parameter sweeps via machine learning could also be used to identify optimal conditions, our analytical approach achieves this in a fraction of the time and permits the calculation of the exact final quantum state for reasonably small cluster states. This allows us to pinpoint the most critical design parameters, facilitating targeted source engineering for higher-fidelity and larger photonic cluster states.

## 2. Methods: Modeling experimental imperfections in the cluster state generation scheme

### 2.1 The Lindner-Rudolph protocol: practical considerations

The Lindner-Rudolph proposal describes an excitation protocol that allows for the generation of arbitrarily long chains of (polarization-)entangled photons in a linear cluster state [15]. This is achieved by a periodically timed excitation of a precessing matter spin, exploiting the transfer of entanglement via optical selection rules [11,25]. In semiconductor quantum dots, the Lindner-Rudolph protocol has been applied to various excitonic complexes of semiconductor quantum dots, such as the dark exciton [11], the positively and the negatively charged trion [12,13,26], and beyond quantum dots it has been applied to trapped atoms [14]. This manuscript considers, for simplicity a quantum dot (QD) trion in Voigt geometry. However, the work presented in this manuscript is not limited to this sub-system and would apply to any other configuration of linear cluster state generation via the Lindner-Rudolph protocol.

The linear cluster state generation via the Lindner-Rudolph protocol is sketched in **Fig. 1.** The creation of a linear cluster state is depicted in a quantum circuit notation (see Fig. 1(a)) by preparing all qubits in the $|+\rangle$ state, for example, by using a Hadamard (H) gate on a $|0\rangle$ state and by subsequently entangling all neighboring qubits by applying controlled Pauli Z gates. Green dots represent the photons entangled through a controlled Z gate.

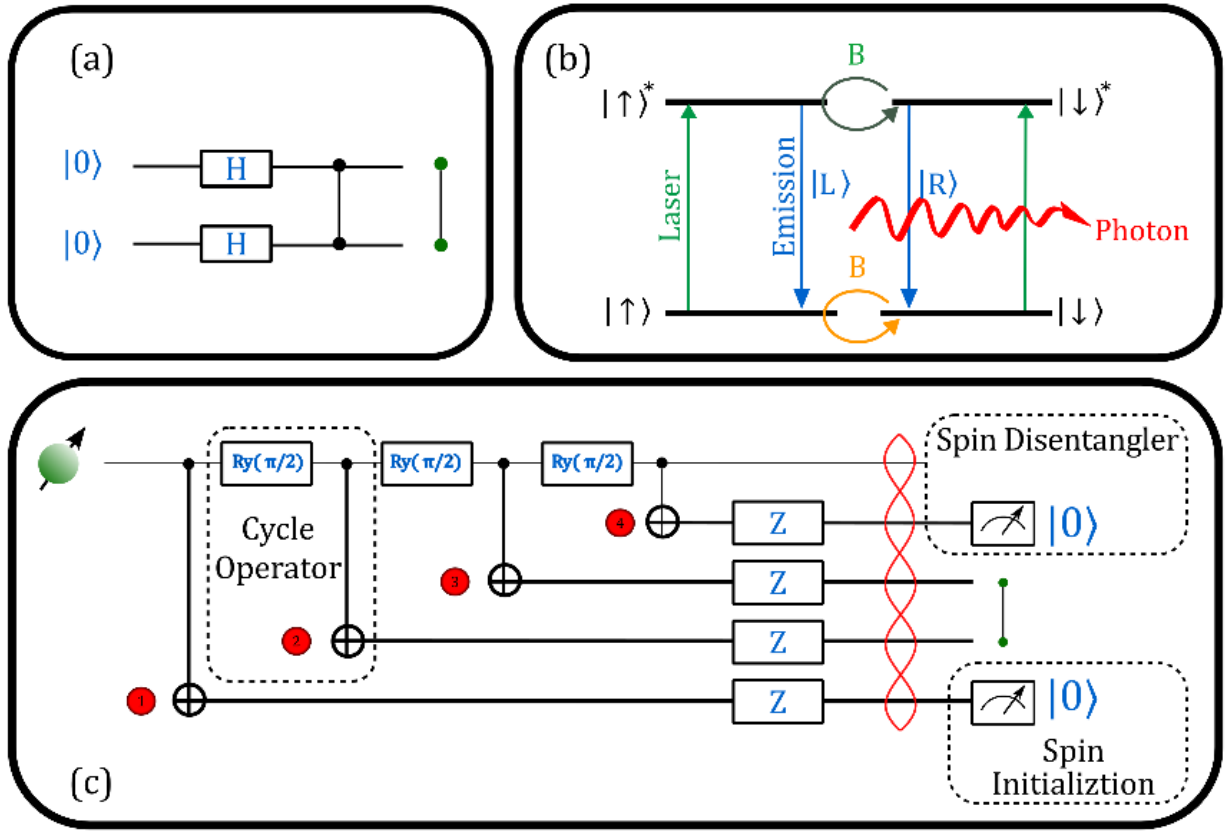

**Fig. 1. Linear cluster states via the Lindner-Rudolph protocol. (a)** Quantum circuit notation. The creation of a linear cluster state can be depicted in a quantum circuit notation by preparing all qubits in the $|+\rangle$ state, e.g., by using a Hadamard (H) gate on a $|0\rangle$ state and entangling all neighboring qubits by applying controlled Pauli Z gates. Green dots represent the photons entangled through a controlled Z gate. **(b)** Schematic of a four-level $\Pi$-system to be used for the cluster state generation. The process begins in the energetic ground state and with a known spin. The system is excited by a laser pulse into an energetically excited state (green arrow). It then relaxes back into the ground state, emitting a polarized photon after a probabilistic lifetime (blue line). This process is polarization-preserving. To control the spin's polarization, we introduce a magnetic field, causing the spin to precess (yellow line) along the y-axis. The spin then also precesses in the excited state, although potentially with an altered precession frequency. **(c)** Full protocol to create a linear cluster state (following Ref.[15]): The spin in the energetic ground state, see (b), can be manipulated and entangled with successively emitted photons. The spin rotates continuously around the y-axis due to the applied magnetic field. Excitation pulses timed at every quarter precession create entangled photons, which are transmitted through waveplates to rotate them by $\pi$ along the z-axis to create a linear cluster state.

A physical way to implement this is, for example, the four-level $\Pi$-system provided by a singly charged exciton (trion) of a semiconductor quantum dot subject to a small magnetic field in Voigt geometry (see Fig. 1(b)).

The Lindner-Rudolph protocol can then be decomposed into the following three steps, essential to our error analysis:

**1. Precession of spin**: Starting with the spin in the $|\uparrow\rangle$ state, the spin precesses around the $(|\uparrow\rangle + i|\downarrow\rangle)$ axis (y-axis of the Poincaré sphere) by applying an external magnetic field in Voigt geometry along the y-axis.

**2. Photon emission**: At each quarter precession, a laser pulse excites the system to an electronically excited state, and its subsequent relaxation results in the emission of a photon while preserving polarization (Fig. 1b). This step is repeated for each additional photon required to build the cluster state.

**3. Photon polarization evolution and spin initialization**: The emitted photons are transmitted through a half-wave plate (HWP) at 0° with respect to the vertical coordinate axis of the laboratory frame to apply a 180° rotation around its $|H\rangle$ axis of Poincaré sphere (taking $|H\rangle$ map to $|0\rangle$), which is equivalent to applying a Z-gate on the matter spin. To initialize and disentangle the spin from the cluster state, the first and the last photons are both projected into the $|\uparrow\rangle$ state via measurement.

The full protocol to create a linear cluster state is then depicted in Fig. 1(c), which includes the initialization and finally the disentangling operation of the matter spin qubit from the photonic qubits.

Having recapitulated the Lindner and Rudolph protocol, we now present a concise survey of the key parameters and physical processes that govern its practical implementation.

A **quasi-resonant excitation** is employed to apply the entangling gate. This can be done, for example, via LO-phonon or LA-phonon assisted excitation. This ensures efficient spectral filtering of the excitation source. Strictly resonant excitation using cross-polarization cannot be applied as the excitation mechanism, as it requires filtering out one polarization basis. Ideally, this excitation occurs instantaneously at a well-defined point in time with perfect efficiency. However, the pump pulse has a finite temporal broadening and a non-unity state preparation fidelity, meaning there is a non-zero probability of failing to excite the system. Please note that the error introduced by a failed excitation is fundamentally different from the error introduced by photon loss [26], even though both will be registered as a non-detection event. Since if the photon is lost, the excitation still alters the spin's precession, while a failure to excite leaves the spin precessing in its ground state.

The **spin precession of the ground state** with a Larmor precession time of $T_{\text{LG}} = 2\pi\hbar/(g_G\mu_B B)$ is dictated by the ground state's Landé g-factor $g_G$ and the external magnetic field $B$. Thus, it defines the timing of the $\frac{\pi}{2}$ rotation gate $R_y$, and in turn the cycle rate $r_c = 4/T_{LG}$ [11, 17]. Consequently, the generation of an $n$-photon cluster state requires $(n + 2)$ quarter-precession periods, inducing a cluster-state generation rate of $r_{\text{LCS}}^n = 4 \cdot \left((n + 2) \cdot T_{\text{LG}}\right)^{-1}$.

The **excited state's lifetime** $\tau$ determines the probabilistic nature of the spin's relaxation back to the ground state. Since this relaxation occurs over a finite timescale and is probabilistic in nature, it introduces an error in the $R_y$ gate rotation. With multiple repetitions, the probabilistic nature of the decay leads to fluctuations in the error, increasing the mixedness of the generated quantum state, requiring $\tau \ll T_L$. The probabilistic decay imposes an upper bound on the clock rate, as a fast clock rate with a long lifetime will produce a completely mixed state. Purcell enhancement can be exploited to reduce the excited state's lifetime $\tau$, thereby reducing the lifetime error.

The **spin precession of the excited state** with period $T_{\text{LE}}$ is dictated by the g-factor of the excitonic complex $g_E$. The excited state g-factor typically differs from that of the ground state, sometimes with an inverted sign. This altered (potentially reversed) precession of the excited state spin, combined with the probabilistic decay of the excited state, affects the mixedness in the $R_y$ gate that varies randomly with each repetition.

**Spin dephasing (decoherence)** arises due to the presence of fluctuating magnetic fields across experimental cycles. A predominant source of these fluctuations is the Overhauser magnetic field induced by the surrounding nuclear spin environment. These variations introduce a phase jitter in the $R_y$ gate, leading to a growing discrepancy between the ideal and experimentally realized spin states as the precession time increases. This accumulated phase uncertainty ultimately results in decoherence and is described by the spin dephasing time $T_2^*$. In the Lindner and Rudolph linear cluster state preparation protocol, the spin is recurrently entangled with each emitted photon. Consequently, the spin undergoes partial re-initialization, which counteracts some of the decoherence effects. The underlying principles and calculations are elaborated upon later in this manuscript.

## 2.2 Formalization and calculation of the gate and state fidelities

To evaluate the impact of the imperfections discussed in an experimental setup on the fidelity of a generated cluster state, we begin by considering an ideal photonic cluster state generated from a spin-based system. Sequentially, we introduce and analyze each source of error. We identify and model the following imperfections: precession errors resulting from 1) the timing of the excitation pulses, 2) decoherence of the spin due to environmental interaction, 3) probabilistic relaxation of spin (lifetime of spin), and 4) difference in angular precession frequency of excited and ground state in the presence of a magnetic field.

In an ideal system, an exact copy of the state is created every time for computation, ensuring it remains pure. However, in an experimental system, imperfections inevitably arise, causing the copies to deviate from being identical and introducing mixedness into the ensemble.

**Fidelity of an ensemble.** Creating a linear cluster state with the Lindner and Rudolph protocol requires repetitive, precise excitation to implement the $R_y\left(\frac{\pi}{2}\right)$ rotation gate on the spin, followed by a controlled-NOT (CNOT) gate that is realized in the photon emission. The accuracy of these gates is crucial for the fidelity of the generated state. Since photon emission occurs via a polarization-preserving deexcitation, and post-selection eliminates events lacking excitation, the fidelity of the CNOT gate is effectively unity. Consequently, the overall fidelity of the cyclic gate sequence is determined solely by the fidelity of the $R_y\left(\frac{\pi}{2}\right)$ gate. Moving forward, we will use gate fidelity to evaluate the quality of the protocol and state fidelity to assess the generated cluster state ensemble. The fidelity of an erroneous operator $O_e$ with respect to an ideal operator $O_I$ is $F_{Operator}(e) = \min_{|\psi\rangle} \text{Fidelity}\left(O_I|\psi\rangle, O_e|\psi\rangle\right)$, The operator's fidelity is defined as the infimum of the fidelity between the ideally and actually evolved states $|\psi\rangle$, taken over all possible input states. For the $R_y$ gate, the state $|\psi\rangle$ equates to $|0\rangle$, simplifying the gate fidelity as $\mathcal{F}_{R_y}(e) = \text{Fidelity}\left(R_y\left(\frac{\pi}{2}\right)|0\rangle, R_y\left(\frac{\pi}{2}+e\right)|0\rangle\right) = \text{Cos}^2\left(\frac{e}{2}\right)$. For an ensemble of experiments with probabilistic errors, we can integrate this fidelity to get the gate fidelity of the protocol, $\mathcal{F}_{R_y} = \int p(e)\cos^2\left(\frac{e}{2}\right)\mathrm{d}e$. Similarly, the ensemble fidelity with ideal cluster state $|\psi_c\rangle$ is defined as $\mathcal{F}_{ensemble} = \langle\psi_c|\rho_{ensemble}|\psi_c\rangle$, where the ensemble state is given by $\rho_{ensemble} = \int p(\rho_e)\,\rho_e\,\mathrm{d}e$, a probabilistic integral of the density matrix $\rho_e$ over parameter space $e$, with the probability distribution density function $p(\rho_e)$. Substituting this into the ensemble fidelity expression, we obtain

$$\begin{aligned}
\mathcal{F}_{ensemble} &= \langle\psi_c|\int p(\rho_e)\rho_e\,\mathrm{d}e|\psi_c\rangle \\
&= \int p(\rho_e)\langle\psi_c|\rho_e|\psi_c\rangle\,\mathrm{d}e \qquad\qquad (1)\\
&= \int p(\mathcal{F}_e)\mathcal{F}_e\,\mathrm{d}e.
\end{aligned}$$

A key outcome of this work is the computation of the fidelity $F_e$. To achieve this, we implement an algorithm that tracks the global density matrix throughout the Lindner and Rudolph protocol. We begin by defining the Hilbert space as the combined Fock space of all optical modes. Each component of the protocol, including the applied gates, the emitted photons, and the error channels, is formulated as a matrix operator acting on this Hilbert space. Using these operators, we analytically derived the post-circuit density matrix $\rho_e$, subsequently used to calculate $F_e$. Moving forward, we introduce errors and parameters progressively while formulating the fidelity of a single cluster state experiment as the function of the imperfection with its probability distributions.

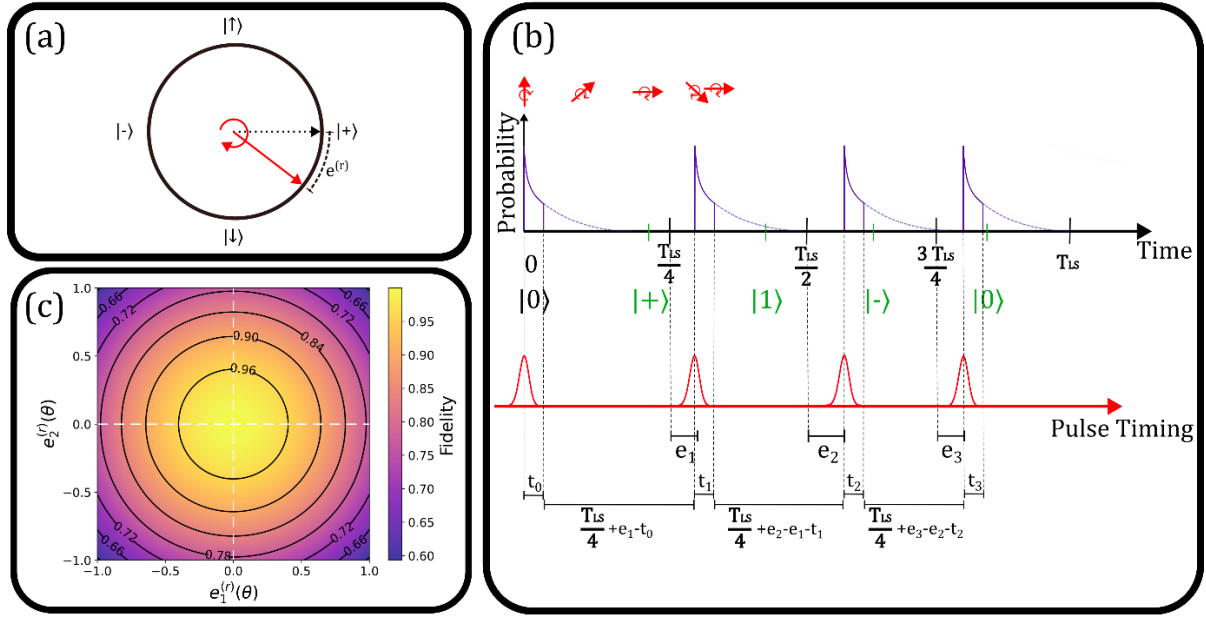

**Fig. 2.** Rotation deviation affecting fidelity. **(a)** The rotation error by the pulse timing is depicted in the Bloch sphere. **(b)** A depiction of the quantum state evolution, incorporating pulse timing errors $e_i$, and $t_i$ denotes the time the spin spends in the excited state during the emission of the $i^{th}$ photon. With the photon emission acting as the partial initialization, the polarization of the $i^{th}$ photon is influenced by the previous relaxation times. The formalization of the effective rotation error has been calculated with $t_i$ as the time of spin in the excited state, an excitation pulse timing error of $e_i$, and an effective angular precession frequency of $\omega'$. With this, the rotation error is $e_i^{(r)} = \left(\frac{\pi}{2\omega} + e_i - e_{i-1} - t_{i-1}\right)\omega' - t_i\omega - \frac{\pi}{2}$. **(c)** A colormap depicting fidelity as a function of the first and second excitation rotation errors, with circular symmetric topography illustrating the independent variables, thereby the localization of errors.

**Pulse timing.** Imperfections in the excitation and relaxation cycle lead to rotation errors $e_i^r$ in the $R_y$ gate at each quarter precession, which in turn affect the polarization of the emitted photon and the fidelity to the target cluster state. When the spin rotation is not driven actively but is a consequence of Larmor precession in a magnetic field, *any deviation of the timing of each excitation pulse* leads to a rotation error, illustrated in **Fig. 2(a)**. Therefore, we define a list of relative excitation time errors $E = \{e_i\}$ where $i \in [0, n+1]$ enumerates the excitation pulse, and $n$ is the number of photons in the cluster state, as illustrated in **Fig. 2(b)** (Since lifetime decay is defined later, the figure can be interpreted with $t_i = 0$, corresponding to instantaneous decay). Since we define the experiment reference time as $t_0 = 0$ s, we set $e_0 = 0$ as well. By leveraging our algorithm to calculate the state fidelity and the gate fidelity for multiple values of $n$, we identified the general formalization of the state fidelity $\mathcal{F}$ and the gate fidelity $\mathcal{F}_{r_y}$ as

$$\mathcal{F}(E^r) = 1 + (-1)^{n+1} \prod_{i=1}^{n+1} \sin\left(e_i^{(r)}\right) + \sum_{1 \leq k \leq \frac{n+1}{2}} \sum_{\substack{J \subseteq \{1,2,\ldots,n+1\} \\ |J|=2k}} \prod_{j \in J} \cos\left(e_j^{(r)}\right) \quad (2a)$$

$$\mathcal{F}_{R_y}\left(e_1^{(r)}\right) = \frac{\cos^2\left(e_1^{(r)}\right)}{2} \quad (2b)$$

where $E^{(r)}(E, \omega)$ is a set of rotation error elements $e_i^{(r)} = (e_i - e_{i-1})\,\omega$, and $\omega$ denotes the angular precession frequency of the spin. Plotted in Fig. 2(c) is the state fidelity graph with varying rotation error $e_i^{(r)}$ caused by the first and the second excitation pulse after the initialization pulse. We observe that the rotation error associated with the $n^{\text{th}}$ photon is independent of the rotation error in the $(n-1)^{\text{th}}$ photon, representing localization of error [15]. This observation suggests that errors in spin polarization do not propagate following the emission of a photon, and an error in the timing of the second photon cannot be compensated by shifting the third photon. Instead, it indicates a form of partial initialization of the spin with each photon emission, effectively resetting the spin state after each excitation event. I.e., irrespective of the deviation in the second pulse timing, the third pulse should be a quarter of the Larmor precession time of the spin precession.

**Decoherence of the spin state.** With the information about excitation timing errors, we move to the second major source of error, that is, *spin decoherence*. As discussed in section 2.1, decoherence can be viewed as a characteristic of a state ensemble, each state exhibiting slightly different angular precession frequencies due to the influence of nuclear spins. While we apply a magnetic field substantially larger than the nuclear Overhauser field to achieve the desired spin precession in accordance with experimental requirements, the component of the nuclear magnetic field aligning with the applied magnetic field significantly impacts the angular precession frequency. This Overhauser field interaction introduces a degree of variability among the angular precession frequency of spins, contributing to the decoherence of the ensemble. Formulating the additional rotation error due to the difference in the effective angular precession frequency of $\omega'$ (arising from the additional nuclear magnetic field) and the excitation pulse timing aligned with the mean angular precession frequency $\omega$, we obtain the updated rotation error $E^{r,2}(E, \omega, \omega')$ with elements $e_i^{r,2}$ defined as

$$e_i^{(r,2)} = \left(\frac{\pi}{2\omega} + e_i - e_{i-1}\right)\omega' - \frac{\pi}{2}. \tag{3}$$

Due to the influence of nuclear spins, the effective angular precession frequencies follow a Gaussian distribution, with the standard deviation of this distribution given by $\sigma_c = \frac{\sqrt{2}}{T_2^*}$, where $T_2^*$ represents the experimentally determined spin dephasing time for a particular sample [18]. This relationship allows us to define the probability distribution function $\mathcal{P}_{\text{frequency}}$ for a state with an angular precession frequency $\omega'$ as

$$\mathcal{P}_{\text{frequency}}\left(\omega', T_2^*\right) = \frac{1}{\sqrt{2\pi\sigma_c^2}} \exp\left(-\frac{(\omega'-\omega)^2}{2\sigma_c^2}\right), \tag{4}$$

where $\sigma_c = \frac{\sqrt{2}}{T_2^*}$.

**Probabilistic decay of the excited state (finite lifetime).** The third considered error, which we will call the *lifetime error*, originates from spin precession in the excited state combined with the probabilistic decay of the excited state. As the spin continues to precess in the excited state, this probabilistic decay leads to rotation error. With the state exhibiting the relaxation time in the excited state, the g-factor of the excited state will govern the precession of the spin. For now, we assume that the g-factor of the excited state is such that the spin precesses in the opposite direction and at the same speed (the variability of the g-factor in the excited state will be explored in the next section). As illustrated in Fig. 2(b) the relaxation time $t_i$, for $i \in [0, n+1]$, corresponds to the expected time it takes for the excited trion state to relax to the ground state. We define $T_d$ as the set of relaxation times for each transition in a single experiment. As the spin starts to precess backward (red arrow in Fig. 2(b)) while in the excited state for time $t_i$, the rotation error set $E^{(r,3)}(E, T_d, \omega, \omega')$ is calculated to be

$$e_i^{(r,3)} = \left(\frac{\pi}{2\omega} + e_i - e_{i-1} - t_{i-1}\right)\omega' - t_i\,\omega' - \frac{\pi}{2}. \tag{5}$$

To formalize the probabilistic distribution function of such a state, we define the state population function $\mathcal{L}$ of the excited states of a quantum dot (or any four-level $\Pi$-system) as an error function (erf) for the rising part, multiplied with an exponentially decaying function, with the lifetime of the excited state as the decay constant. This read

$$\mathcal{L}(t, \tau_r, \tau_d) = \frac{1}{2}\left(1 + \text{erf}\left(\frac{t}{\sqrt{2\tau_r}}\right)\right)\exp\left(-\frac{t}{\tau_d}\right), \qquad (6)$$

where $\tau_r$ is the rising constant and $\tau_d$ is the lifetime of the excited state population, which is characteristic for each transition line of a quantum dot device and can be determined experimentally. To derive the probabilistic distribution function $P_{\text{lifetime}}$ for the cluster state, the joint probability is the product of the individual probabilistic population functions, expressed as

$$\mathcal{P}_{\text{lifetime}}(T_r, \sigma_l, T_d) = \prod_{i=0}^{n+1} \mathcal{L}(t_i, \tau_r, \tau_d). \qquad (7)$$

**Excited state and ground state g-factor.** We consider constant g-factors, treating them as scalars instead of tensors. This simplification is appropriate within the scope of this manuscript. For a detailed study on the impact of the g-factor anisotropy on spin-photon entanglement generation, see Ref. [27]. For simplicity in the formulation of the state population function, we initially assumed that the angular precession frequency of the spin in the excited state was equal and opposite to the ground state. However, the angular precession frequency and direction may vary depending on the Landé g-factor of the excited state. To account for this variation, we introduce $g_{\text{ratio}} = \frac{g_{\text{excited}}}{g_{\text{ground}}}$ into the equation, defined as the ratio of the Landé g-factor for the excited state to that of the ground state. The precession of the spin in the excited state for $\tau_d$ time adds an additional factor onto the rotation error caused by the probabilistic lifetime decay. This gives us the rotation error set $E^{(r,4)}(E, T_r, \omega, \omega', g_{\text{ratio}})$,

$$e_i^{(r,4)} = \left(\frac{\pi}{2\omega} + e_i - e_{i-1} - t_{i-1}\right)\omega' + g_{\text{ratio}}\, t_i\, \omega' - \frac{\pi}{2}.\tag{8}$$

As the probability of a certain angular precession frequency $\omega'$ and probabilistic decay of excited state are independent, we can determine the joint probability for an individual cluster state by taking the product of the two probabilities. This gives the probability of the state with error $E^{(r,4)}(E, T_r, \omega, \omega', g_{\text{ratio}})$ as

$$\mathcal{P}(T_r, \sigma_l, \tau_d, \omega', T_2^*) = \mathcal{P}_{\text{frequency}}(\omega', T_2^*) \cdot \mathcal{P}_{\text{lifetime}}(T_r, \sigma_l, \tau_d).\tag{9}$$

With the fidelity of an individual cluster state $\mathcal{F}(E^{(r,4)}(E, T_r, \omega, \omega', g_{\text{ratio}}))$ and its probability distribution function $P$, we can integrate over the probabilistic distribution to get the fidelity of the ensemble, i.e., the experimentally measured state measuring multiple copies of the same single state,

$$\mathcal{F}_{\text{ensemble}}(E, \omega, g_{ratio}, \sigma_l, \tau_d, T_2^*) = \iiint_{-\infty}^{\infty} \mathcal{F}\left(E^{(r,4)}(E, T_r, \omega, \omega', g_{ratio})\right)$$
$$\cdot\, \mathcal{P}(T_r, \sigma_l, \tau_d, \omega', T_2^*)\, d\omega'\, dt_0 dt_1 \dots dt_{n+1},\tag{10}$$

$$\mathcal{F}_{R_y}(e_0, e_1, \omega, g_{\text{ratio}}, \sigma_l, \tau_d, T_2^*) = \iiint_{-\infty}^{\infty} \cos^2\left(\frac{e_1^{(r,4)}}{2}\right) \cdot \mathcal{P}(T_r, \sigma_l, \tau_d, \omega', T_2^*)\, d\omega'\, dt_0\, dt_1.\tag{11}$$

To simplify the analytical integration, we assume that the excitation of the system can be performed infinitely fast, with $\tau_r$ to 0, and replace the error function in the probability distribution function describing the population $\mathcal{L}$ with a step function. This simplifies the integration for $t_i$ to be performed with the limit of $[0, \infty)$. In certain experiments, a temporal post-selection is applied to the lifetime decay, which serves to reduce the state's mixedness and thereby enhance the overall fidelity. For such cases, the fidelity can be computed by integrating the eq. (10) over a restricted time window of $[0, T_{bin}]$ to reflect the experimental selection criteria. Using a computer algebra system to integrate eq. (10) and eq. (11), in our case Wolfram *Mathematica*, we derive the analytical solution for the ensemble fidelity, $F_{\text{ensemble}}$ as well as the fidelity, $F_{R_y}$ of the $i^{\text{th}}$ $R_y$ gate applied to the spin. The analytical solution can, in principle, be found for any larger linear cluster state. Solving the integration becomes an $n$-fold exponentially growing problem. Thus, we analyze relatively small states of 2-4 qubits, but any conclusion from these small systems can readily be extended to states of arbitrary size.

## 3. Results and Discussion

In this section, we use the mathematical model described in Section 2.2 to evaluate the impact of various physical parameters on the gate fidelity of the entangling operation and on the state fidelity of the created $n$-qubit LCS.

In **Fig. 3**, we focus on the fidelity of the entangling gate and evaluate the impact of the excited state lifetime, the spin coherence time, and the ground and excited state g-factors. We then examine how these parameters collectively influence the fidelity when considered together. We identify an optimal angular precession frequency of the spin, as well as excitation timing, for a realistic physical system. If not otherwise specified, it is natural to use the Larmor precession period of the ground state $T_{\text{LG}}$ as the unit of time in our figures. **Fig. 3(a)** presents the gate fidelity as a function of the excited-state lifetime in the absence of spin decoherence. A shorter excited-state lifetime leads to a narrower distribution of the rotation error $R_y(\pi + e^{(r)})$, effectively reducing the degree of mixedness and enhancing the gate fidelity, and we find that $F_{R_y} > 0.99$ for $\tau < 0.03 \cdot T_{\text{LG}}$.

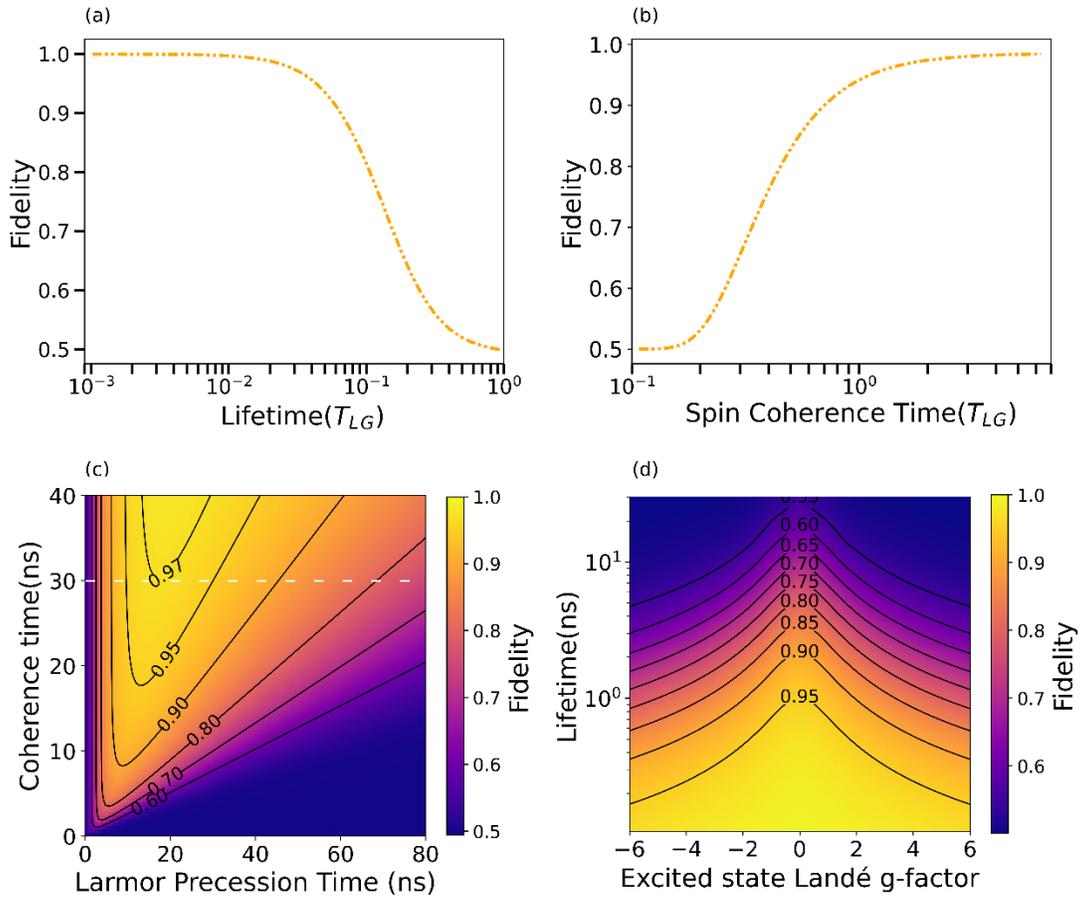

**Fig. 3.** Parameters determining the $R_y(\frac{\pi}{2})$ gate fidelity. **(a)** Lifetime error. **(b)** Spin decoherence. **(c)** False-color plot of gate fidelity as a function of the spin Larmor precession time and coherence time, assuming a fixed radiative lifetime of $\tau = 400$ ps. The plot highlights two dominant sources of mixedness: lifetime-induced decay at short precession times and spin dephasing at long precession times. An optimal precession time emerges where these effects are balanced, yielding maximal gate fidelity for given coherence and lifetime values. **(d)** Gate fidelity as a function of the excited state Landé g-factor normalized to the ground state Landé g-factor.

Similarly, **Fig. 3(b)** shows the gate fidelity as a function of the spin coherence time, assuming instantaneous decay from the excited state. With coherence times shorter than the quarter-period of Larmor precession, $T_2^* < \frac{1}{4} \cdot T_{\mathrm{LG}}$, each quarter precession effectively results in a fully mixed spin state. This is reflected by the gate fidelity dropping to 0.5, indicating a complete loss of quantum coherence. As $T_2^*$ increases, the gate fidelity increases and approaches unity, with $F_{R_y} >$ 0.99 for $T_2^* > 1.8\,T_{LG}$. To explore the interplay between lifetime, angular precession frequency, and decoherence, **Fig. 3(c)** presents the gate fidelity for a fixed excited-state lifetime of 400 ps as a function of the Larmor precession time and spin coherence time.

As shown in Fig. 3(a), in the absence of decoherence, a slower spin precession is advantageous; however, Fig. 3(b) illustrates that in the presence of decoherence, a slower precessing spin becomes less suitable. This highlights a *fundamental trade-off between decoherence and lifetime*, which must be carefully balanced to optimize the cluster state generation. Fig 3(c) illustrates the existence of an optimal Larmor precession time, which increases with longer spin coherence times. Extended coherence not only shifts this optimum but also allows greater flexibility in the timing of spin manipulations, enabling high-fidelity cluster state generation over a broader temporal range.

Since the spin angular frequencies of the ground and excited states can differ, Fig. 3(d) presents the gate fidelity as a function of the g-factor ratio (excited state to the ground state) and lifetime. The coherence time is set to 30 ns, and the optimal Larmor precession time is taken for each corresponding lifetime and coherence time shown in the plot.

Since the Landé g-factor directly influences the precession rate, it plays a critical role in determining the mixedness introduced by the probabilistic decay of the excited state. A slower precession in the excited state, corresponding to a smaller g-factor, reduces phase uncertainty during decay and thus preserves the purity of the spin state.

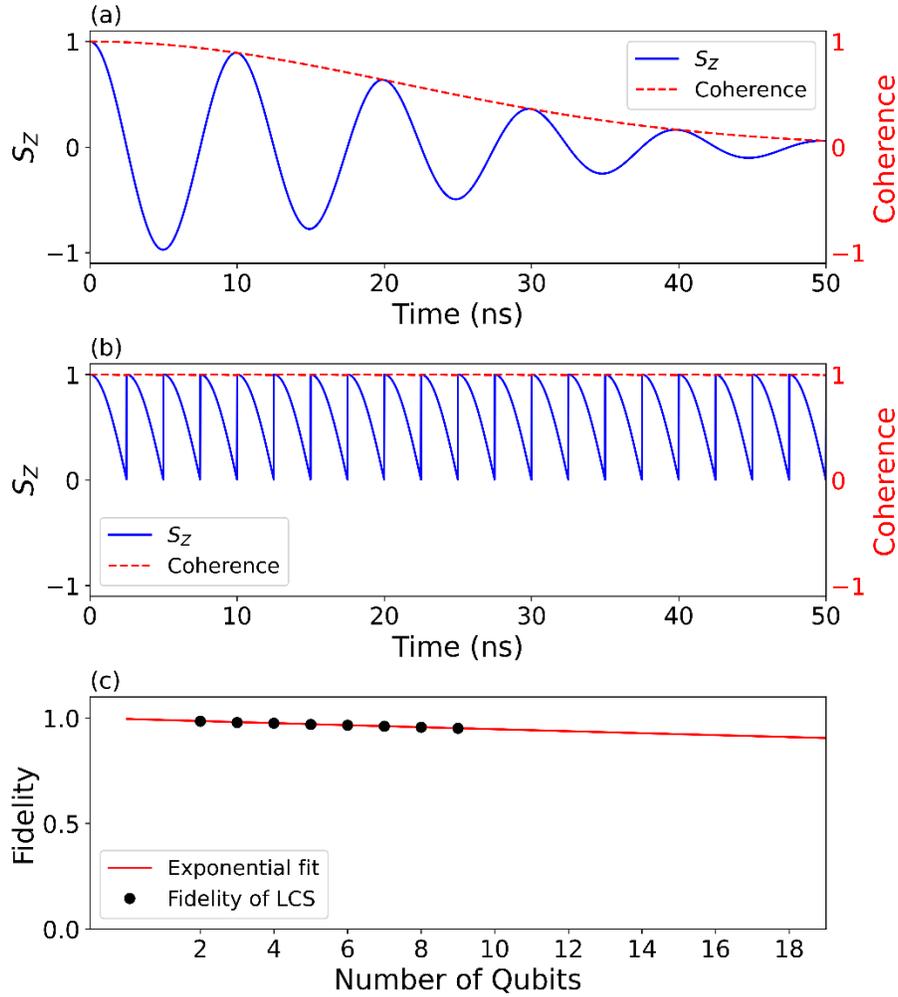

**Fig. 4.** Spin decoherence and partial re-initialization. **(a)** Spin polarization time trace (blue line) for $T_2^* = 30$ ns and a Larmor precession time of $T_{LG} = 10$ ns. The decay envelope (red dashed line) represents the coherence of the spin. **(b)** Illustration of the partial spin re-initialization through photon emission caused by localization of error, maintaining the coherence of the system. **(c)** The time axis of the $n$-qubit cluster state is scaled to match the emission timing of the $(n + 2)^{\text{th}}$ photon during spin precession. This accounts for the $n$ photons forming the linear cluster state, along with one initialization photon and one disentangling photon. By aligning the time axis in this way, we demonstrate that the spin can sustain high-fidelity cluster state generation even in the presence of a short decoherence time, $T_2^*$.

In the ideal case, a spin system with effectively "frozen" precession in the excited state minimizes the mixedness associated with the emission of the $n^{\text{th}}$ photon. This is the case when the excited state g-factor is zero. However, the time spent in the excited state still shifts to the quarter-precession point of the subsequent $(n + 1)^{\text{th}}$ photon where it localizes as error. This interplay highlights the importance of maintaining both a low excited-state g-factor (relative to the ground state) and, more importantly, a short radiative lifetime to achieve high-fidelity cluster state generation.

The next property we analyze is the partial initialization of the matter spin decoherence affecting the fidelity of the linear cluster state. In **Fig. 4 (a)**, we plot the third Stokes parameter ($S_Z$) of the matter spin, representing the precession and coherence ($\sqrt{S_z^2 + S_y^2 + S_x^2}$) of the spin over time since spin initialization. Exemplarily, a spin coherence time of $T_2^* = 30$ ns is chosen, and $T_{\text{LG}} = 10$ ns. F**ig. 4 (b)** illustrates the partial spin reinitialization occurring with each photon emission, which helps maintain the spin's coherence. Finally, **Fig. 4 (c)** shows the fidelity of an $n$-qubit linear cluster state, with the axis scaled to align with the emission timing of the $n^{th}$ photon during precession. For the fidelity calculations, the lifetime is zero, the precession period is assumed to be 10 ns, and the spin dephasing time is 30 ns. In Fig. 4 (c), we observe an exponentially decaying fidelity as the cluster state length increases. However, this decay rate is notably smaller than that of the decoherence; this reduced decay is attributed to partial spin initialization. These results suggest that a cluster state with a construction time exceeding the system's coherence time can be generated. However, with all other parameters fixed, a larger coherence time significantly improves the cluster state fidelity, as illustrated in Fig. 4 (b).

One of the key effects of partial reinitialization is demonstrated in **Fig. 5**, where both the entangling gate fidelity and the cluster state fidelity are plotted as functions of the Larmor precession time for a fixed coherence time of 30 ns and a lifetime of 400 ps. Each photon emission acts as a partial reinitialization of the spin state, effectively localizing spin errors and decoupling the $R_y$ gate operations from one another. As a result, the optimal Larmor precession time becomes identical for all entangling gates, remaining constant regardless of the cluster state length. Hence, the optimal operational parameters, once identified for the entangling gate, can be used for the generation of any $n$-qubit LCS.

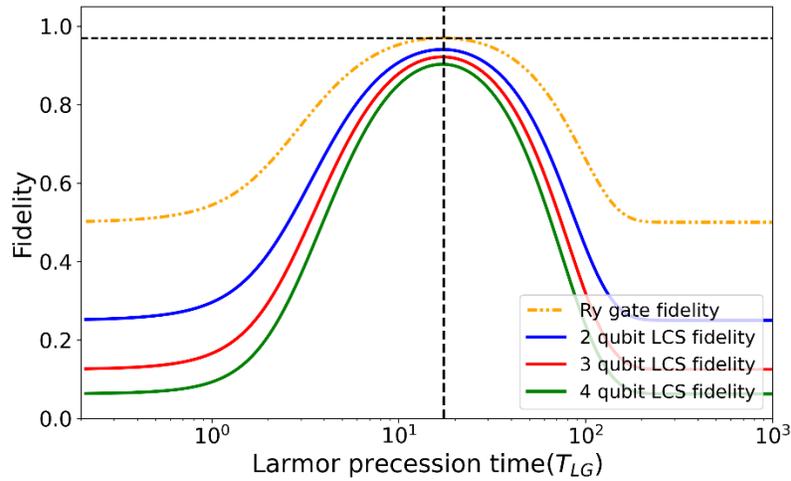

**Fig. 5.** Entangling gate and cluster-state fidelity as a function of spin Larmor precession time. The entangling gate fidelity and linear cluster-state fidelity for 2-, 3-, and 4-qubit chains, plotted as a function of precession time for a fixed coherence time of 30 ns (corresponding to the white dashed line in Fig. 3 (c)). The curves reveal a consistent optimal precession time $T_{LG}$, invariant with respect to cluster length due to partial spin reinitialization.

We now incorporate all sources of error into the system, enabling us to identify the optimal tunable parameters that yield the highest fidelity. Working on a system with a lifetime of 400 ps, decoherence of 30 ns, and an excited state g-factor of -3, the optimal Larmor precession time came out to be 14 ns. With this, we optimized the pulse timing for forming a 2-qubit cluster state.

The finite lifetime and different g-factors for excited and ground state spin necessitate a correction factor on the excitation timing. This adjustment ensures that the average deexcitation probability remains close to a quarter precession.

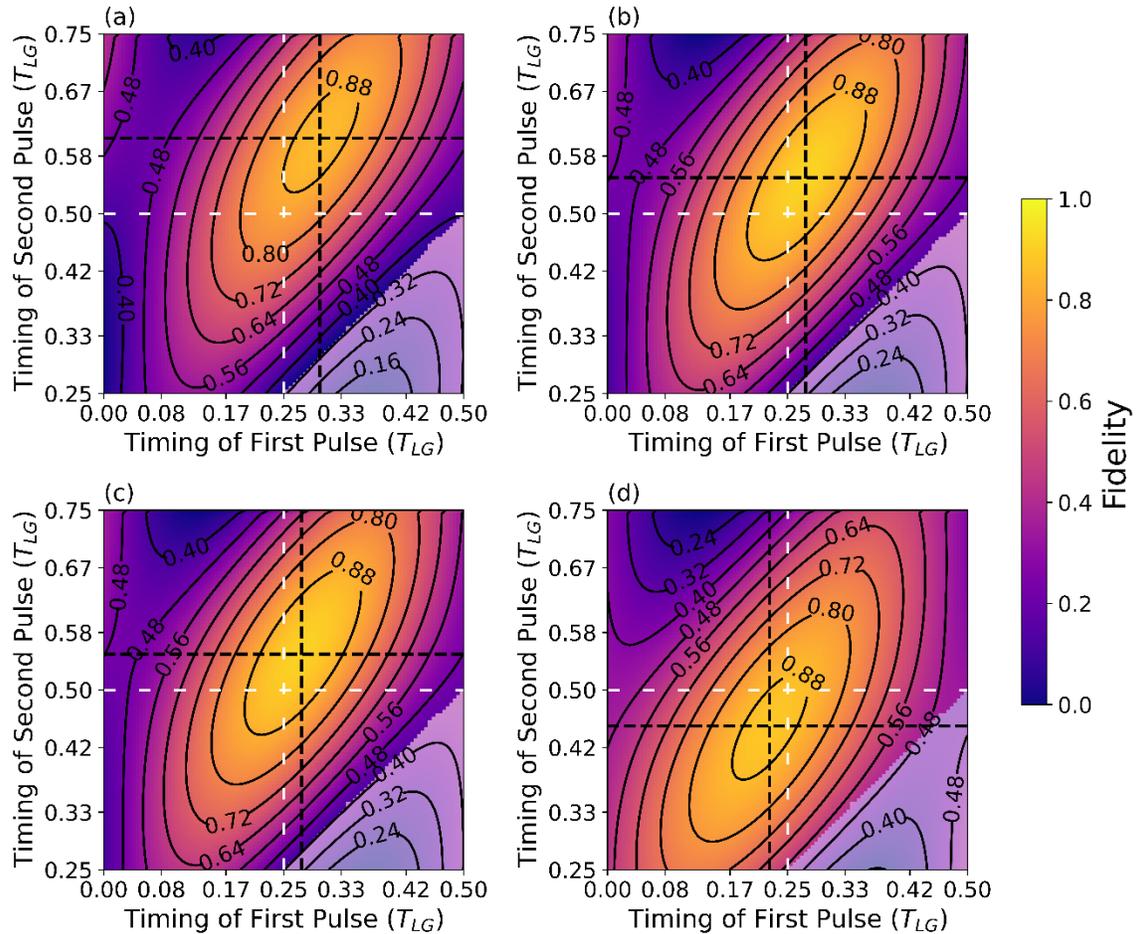

**Fig. 6.** Excitation timing correction. Fidelity of a 2-photon cluster state as a function of the first and second excitation timing. **(a)** For a system with a lifetime of $\tau = 400$ ps, spin coherence time of $T_2^* = 30$ ns, and $T_{LG} = 14$ ns, excited-to-ground state g-factor ratio of $-3$. The maximum fidelity shifts, resulting in a cycle time of $0.30\,T_L$ instead of $1/4 \cdot T_L$. **(b)** Parameters as in (a), but with $\tau = 200$ ps. **(c)** Parameters as in (a) but with an excited-to-ground state g-factor ratio of $-1$. The shift shortens due to the threefold reduction in the excited state precession speed, **(d)** and if the excited-state g-factor is set to 3, causing the spin to precess in the same direction in the excited state, the optimal pulse timing crosses 0.25 and 0.5 marks for the first and second pulse timing, respectively, to compensate for probabilistic decay of the excited state.

Fig. 6(a) presents a false-color plot of the two-qubit linear cluster state (LCS) fidelity as a function of the excitation pulse timing, normalized to the ground-state Larmor precession period $T_{LG}$. The optimal excitation cycle occurs at approximately $T_{LG} = 0.30$, where the fidelity is maximized. In Fig. 6(b), the system parameters are taken to be the same as in Fig. 6(a), except the radiative lifetime is reduced by half. This results in reduced spin precession during the excited-state evolution, thereby shifting the fidelity maxima to $0.275\,T_{LG}$ cycle time. Fig. 6(c) explores a scenario where the $g$-factor ratio between the excited and ground states is reduced by half, to -1. The slower precession in the excited state leads to a shorter shift of $0.275\,T_{LG}$. In contrast, Fig. 6(d) considers a $g$-factor ratio of 3; in this case, the spin continues to precess in the same direction as in the ground state. To compensate, the excitation pulse must be sent earlier in the cycle to maintain high fidelity, as reflected by the shifted fidelity maximum to $0.225\,T_{LG}$ cycle time.

In **Table 1**, we use our method to provide a survey on the fidelities that could be reached with current semiconductor quantum dot technology. Four different subsystems are considered, for which we calculate an upper bound on the $R_y$ gate fidelity as well as the state fidelity of a 3-photon LCS and a 7-photon LCS. Four scenarios are considered: **(i)** The combined state-of-the-art of GaAs QDs. **(ii)** The QD linear cluster state source, which has been used to demonstrate continuous operation [26], and entanglement up to 10 photons [12]. **(iii)** Quantum dot single-photon sources emitting in the telecom C-band fabricated in our laboratory. **(iv)** A QD single-photon source employing the negatively charged trion as an entangler [13]. The gate fidelity $F_{R_y}$ is calculated as it provides the upper bound for the entangling operation. Whereas the state fidelity for a 3-photon

LCS and a 7-photon LCS is provided, as these states are of relevance to quantum computing and communication protocols. For example, a 6-ring cluster state —useful for constructing the fusion network in fusion-based quantum computing— can be generated from a 7-photon LCS via type-I fusion.

| | State-of-the-Art GaAs QDs | QD LCS [12] | Telecom C-Band QDs | QD neg. trion [13] |
|---|---|---|---|---|
| Excited state lifetime $\tau$ | 23 ps [28] | 398 ps [31] | 180 ps [32] | 200 ps [13] |
| Spin coherence time $T_2^*$ | 535 ns (hole) [29] | 30 ns (hole) [33] | 20 ns (hole) | $\approx 2$ ns |
| Landé factor ratio $g_{ex}/g_{gs}$ | $\approx 0$ [30] | $0.37/-0.12 = -3.0$ [31] | $\lvert 2.14 \rvert / \lvert 0.37 \rvert = 5.78$ | $\approx -1/3$ |
| Clock rate, $\omega_C$ | 1.0 GHz | 456 MHz [26] | 212 MHz | 1.23 GHz [13] |
| Comment | Combined SotA | w/o temporal post-selection | | $T_2^*$ specified as *'a few ns'* |
| | | | | g-factor ratio taken from [31] |
| $R_y$ gate fidelity | 0.99965 | 0.86206 | 0.94620 | 0.89385 |
| 3-photon LCS fidelity | 0.99869 | 0.52891 | 0.80157 | 0.63842 |
| 7-photon LCS fidelity | 0.99739 | 0.30567 | 0.64251 | 0.40749 |

Table 1: Comparison of the upper bounds on fidelity across different quantum dot material systems, based on their experimentally determined parameters.

The relevant parameters of the combined state-of-the-art of GaAs QDs are taken from different publications. The Purcell-enhanced lifetime of $\tau = 23$ ps was measured for an excitonic decay of a QD integrated into a circular Bragg grating resonator [28], while the heavy-hole spin coherence time of $T_2^*$ was determined in an open-cavity setup [29]. The trion's excited-state electron g-factor can be tuned to zero by choosing appropriate growth conditions [30]. Though measured separately, these parameters can feasibly be combined in a single device without conflict. This suggests that near-unity gate and state fidelities are achievable at a clock cycle of 1 GHz, and that material properties are no longer the limiting factor. Errors from the excitation scheme, such as pulse width, re-excitation, and state preparation fidelity, may now dominate.

In the second column, we present a quantum dot (QD) system that has thus far demonstrated the longest localizable entanglement length, extending over 10 photons: a positively charged trion in an InAs/GaAs QD, utilizing the heavy-hole spin [23]. Due to low fidelity in the experimental results, temporal post-selection based on the emitter's lifetime was applied to reduce the state's mixedness and enhance the measured entanglement quality.

The third column highlights a QD system emitting in the telecom C-band, which is particularly promising for applications in quantum communication and distributed quantum computing due to the low-loss transmission achievable in standard optical fiber networks. The parameters considered in the calculation are yet unpublished experimental values measured in our laboratory. Intensive ongoing research on the advancement of these sources (for example crystal growth, deterministic dot placement, and electrical tuning) promises further progress. These fidelities can be reached for an optimized clock rate of 212 MHz, which corresponds to an external magnetic field strength of $B \approx 10$ mT. To achieve even higher fidelities close to unity, and to enable higher clock rates, the lifetime needs to be further improved, for example via the Purcell effect. The theoretical model offers early insights into the optimal combination of material and fabrication parameters to enhance tunability and performance.

The fourth column discusses a QD system based on a negatively charged trion, which has demonstrated spin–photon–photon entanglement at the highest clock cycle achieved to date. For our calculations, we considered $T_2^* = 2$ ns, as in Ref. [13] it is specified as *'a few ns'*. Notably, this system exhibits a Larmor precession time that is comparable to or exceeds the decoherence time yet still maintains observable entanglement. This highlights the presence of a phenomenon that sustains the overall coherence of the system during the creation of the cluster state, in contrast to the lower coherence values typically observed through conventional coherence time measurements.

## 4. Conclusions

In this work, we introduced an analytical framework for guiding experimental design and enhancing the system performance across diverse conditions in the context of photonic cluster state generation from quantum dot single-photon sources. Our results indicate that system fidelity is governed by a complex interplay of multiple factors. For a fixed sample characterized by specific lifetime, coherence time, and g-factor values, we determined the optimal fidelity achievable at a well-defined angular precession frequency. Moreover, excitation pulses must be adjusted linearly to counteract the probabilistic decay associated with the finite lifetime. Notably, for a given sample, the optimal angular precession frequency and pulse shift remain unchanged across different chain lengths due to the presence of partial reinitialization. Since the role of partial spin reinitialization in preserving coherence has not been explored in this study, we encourage any experimental endeavor on that matter. On a more fundamental level, our calculations demonstrate a trade-off between the spin coherence, cycle time and excited state lifetime, which must be carefully balanced to optimize the fidelity of the final photonic cluster state. In fact, we found that for any values of spin coherence and lifetime, there is a maximal entangling gate fidelity. We are confident that our model and analytical framework, ant its results derived here, will be useful fur optimizing future experiments for building photonic cluster states.

## Declaration

**Acknowledgements.** We would like to acknowledge Prof. Dr. Jonathan Finley for the insightful discussion of error localization and decoherence of the electron spin in semiconductor quantum dots.

**Availability of data and material.** The datasets used and/or analysed during the current study are available from the corresponding author on reasonable request.

**Competing interests.** The authors declare that they have no competing interests.

**Authors' Information.** No information provided.

**Funding.** The authors acknowledge the support of the state of Bavaria and the German Ministry for Research and Education (BMBF) within Project PhotonQ (FKZ: 13N15759 and FKZ: 13N15761), and Qecs (FKZ: 13N16272).

**Authors' contributions.** S. H. initiated the work. R.P. and S.R. formulated and analyzed the protocol with supervision from P.v.L., incorporating theoretical and experimental insights provided by G.P.. Y.R. contributed to interpreting and illustrating the physical model and proofreading the manuscript. The development of the protocol and the manuscript was carried out under the guidance of T.H.-L. and A.T.P., who provided critical revisions and refinements. All authors read and approved of the final manuscript.

# References

[1] E. Knill, R. Laflamme, and G. J. Milburn, A scheme for efficient quantum computation with linear optics, Nature **409**, 46 (2001). DOI: 10.1038/35051009

[2] R. Raussendorf and H. J. Briegel, A One-Way Quantum Computer, Phys. Rev. Lett. **86**, 5188 (2001). DOI: 10.1103/PhysRevLett.86.5188

[3] H. J. Briegel and R. Raussendorf, Persistent Entanglement in Arrays of Interacting Particles, Phys. Rev. Lett. **86**, 910 (2001). DOI: 10.1103/PhysRevLett.86.910

[4] P. Walther, K. J. Resch, T. Rudolph, E. Schenck, H. Weinfurter, V. Vedral, M. Aspelmeyer, and A. Zeilinger, Experimental one-way quantum computing, Nature **434**, 169 (2005). DOI: 10.1038/nature03347

[5] S. Bartolucci et al., Fusion-based quantum computation, Nat. Commun. **14**, 912 (2023). DOI: 10.1038/s41467-023-36493-1

[6] A. Melkozerov, A. Avanesov, I. Dyakonov, and S. Straupe, Analysis of optical loss thresholds in the fusion-based quantum computing architecture, APL Quantum **1**, (2024). DOI: 10.1063/5.0214728

[7] D. Istrati et al., Sequential generation of linear cluster states from a single photon emitter, Nat. Commun. **11**, 5501 (2020). DOI: 10.1038/s41467-020-19341-4

[8] S. Chen et al., Heralded Three-Photon Entanglement from a Single-Photon Source on a Photonic Chip, Phys. Rev. Lett. **132**, 130603 (2024). DOI: 10.1103/PhysRevLett.132.130603

[9] H. Cao, L. M. Hansen, F. Giorgino, L. Carosini, P. Zahálka, F. Zilk, J. C. Loredo, and P. Walther, Photonic Source of Heralded Greenberger-Horne-Zeilinger States, Phys. Rev. Lett. **132**, 130604 (2024). DOI: 10.1103/PhysRevLett.132.130604

[10] N. Kiesel, C. Schmid, U. Weber, G. Tóth, O. Gühne, R. Ursin, and H. Weinfurter, Experimental Analysis of a Four-Qubit Photon Cluster State, Phys. Rev. Lett. **95**, 210502 (2005). DOI: 10.1103/PhysRevLett.95.210502

[11] I. Schwartz, D. Cogan, E. R. Schmidgall, Y. Don, L. Gantz, O. Kenneth, N. H. Lindner, and D. Gershoni, Deterministic generation of a cluster state of entangled photons, Science (80-. ). **354**, 434 (2016). DOI: 10.1126/science.aah4758

[12] D. Cogan, Z.-E. Su, O. Kenneth, and D. Gershoni, Deterministic generation of indistinguishable photons in a cluster state, Nat. Photonics **354**, 434 (2023). DOI: 10.1038/s41566-022-01152-2

[13] N. Coste et al., High-rate entanglement between a semiconductor spin and indistinguishable photons, Nat. Photonics **17**, 582 (2023). DOI: 10.1038/s41566-023-01186-0

[14] P. Thomas, L. Ruscio, O. Morin, and G. Rempe, Efficient generation of entangled multiphoton graph states from a single atom, Nature **608**, 677 (2022). DOI: 10.1038/s41586-022-04987-5

[15] N. H. Lindner and T. Rudolph, Proposal for Pulsed On-Demand Sources of Photonic Cluster State Strings, Phys. Rev. Lett. **103**, 113602 (2009). DOI: 10.1103/PhysRevLett.103.113602

[16] Y. Reum, Manuscript in preparation: Fault-tolerant optical quantum computing is closer than you might think — thanks to semiconductor quantum dots, (2025). DOI: Not available (manuscript in preparation)

[17] S. C. Wein, T. G. de Brugière, L. Music, P. Senellart, B. Bourdoncle, and S. Mansfield, Minimizing resource overhead in fusion-based quantum computation using hybrid spin-photon devices, (2024). DOI: 10.48550/arXiv.2412.08611

[18] M. L. Chan, T. J. Bell, L. A. Pettersson, S. X. Chen, P. Yard, A. S. Sørensen, and S. Paesani, Tailoring Fusion-Based Photonic Quantum Computing Schemes to Quantum Emitters, PRX Quantum **6**, 020304 (2025). DOI: 10.1103/PRXQuantum.6.020304

[19] F. Wiesner, H. M. Chrzanowski, G. Pieplow, T. Schröder, A. Pappa, and J. Wolters, The influence of experimental imperfections on photonic GHZ state generation, New J. Phys. **26**, 113021 (2024). DOI 10.1088/1367-2630/ad916f

[20] Z. Qin, W.-R. Lee, B. DeMarco, B. Gadway, S. Kotochigova, and V. W. Scarola, Quantifying entanglement in cluster states built with error-prone interactions, Phys. Rev. Res. **3**, 043118 (2021). DOI: 10.1103/PhysRevResearch.3.043118

[21] R. Raussendorf, J. Harrington, and K. Goyal, A fault-tolerant one-way quantum computer, Ann. Phys. (N. Y). **321**, 2242 (2006). DOI: 10.1016/j.aop.2006.01.012.

[22] C. M. Dawson, H. L. Haselgrove, and M. A. Nielsen, Noise Thresholds for Optical Quantum Computers, Phys. Rev. Lett. **96**, 020501 (2006). DOI: 10.1103/PhysRevLett.96.020501

[23] P. Aliferis and D. W. Leung, Simple proof of fault tolerance in the graph-state model, Phys. Rev. A **73**, 032308 (2006). DOI: 10.1103/PhysRevA.73.032308

[24]    S. Brandhofer, I. Polian, S. Barz, and D. Bhatti, Hardware-efficient preparation of architecture-specific graph states on near-term quantum computers, Sci. Rep. **15**, 2095 (2025). DOI: 10.1038/s41598-024-82715-x

[25]    M. Titov, B. Trauzettel, B. Michaelis, and C. W. J. Beenakker, Transfer of entanglement from electrons to photons by optical selection rules, New J. Phys. **7**, 186 (2005). DOI: 10.1088/1367-2630/7/1/186

[26]    Z.-E. Su, B. Taitler, I. Schwartz, D. Cogan, I. Nassar, O. Kenneth, N. H. Lindner, and D. Gershoni, Continuous and deterministic all-photonic cluster state of indistinguishable photons, Reports Prog. Phys. **87**, 077601 (2024). DOI: 10.1088/1361-6633/ad4c93

[27]    P. R. Ramesh et al., The impact of hole $g$-factor anisotropy on spin-photon entanglement generation with InGaAs quantum dots, 1 (2025). DOI: 10.48550/arXiv.2502.07627

[28]    M. B. Rota et al., A source of entangled photons based on a cavity-enhanced and strain-tuned GaAs quantum dot, ELight **4**, 13 (2024). DOI: 10.1186/s43593-024-00072-8

[29]    M. Hogg et al., Fast optical control of a coherent hole spin in a microcavity, 1 (2024). DOI: 10.48550/arXiv.2407.18876

[30]    C. Schimpf et al., Optical and magnetic response by design in GaAs quantum dots, (2025). DOI: 10.48550/arXiv.2504.02355

[31]    D. Cogan, Z.-E. Su, O. Kenneth, and D. Gershoni, Spin purity of the quantum dot confined electron and hole in an external magnetic field, Phys. Rev. B **105**, L041407 (2022). DOI: 10.1103/PhysRevB.105.L041407

[32]    J. Kaupp, Y. Reum, F. Kohr, J. Michl, Q. Buchinger, A. Wolf, G. Peniakov, T. Huber-Loyola, A. Pfenning, and S. Höfling, Purcell-Enhanced Single-Photon Emission in the Telecom C-Band, Adv. Quantum Technol. (2023). DOI: 10.1002/qute.202300242

[33]    D. Cogan, O. Kenneth, N. H. Lindner, G. Peniakov, C. Hopfmann, D. Dalacu, P. J. Poole, P. Hawrylak, and D. Gershoni, Depolarization of Electronic Spin Qubits Confined in Semiconductor Quantum Dots, Phys. Rev. X **8**, 041050 (2018). DOI: 10.1103/PhysRevX.8.041050